\pdfoutput=1

\documentclass[11pt]{article}

\usepackage[textsize=small,textwidth=21mm]{todonotes}
\usepackage{fullpage}
\usepackage{tikz}
\usetikzlibrary{patterns}
\usepackage{subfigure}
\usepackage{xspace}
\usepackage{amsmath}
\usepackage{amsfonts}
\usepackage{color}
\usepackage{float}
\usepackage{amsthm}
\usepackage{amssymb}
\let\chapter\section
\usepackage[ruled,vlined]{algorithm2e}
\usepackage{mathdots}
\usepackage{url}
\usepackage{footnote}
\usepackage[bottom]{footmisc}
\usepackage{tablefootnote}
\definecolor{orange}{rgb}{1,0.5,0}
\definecolor{darkblue}{rgb}{0,0,0.8}
\usepackage[bookmarks=false]{hyperref}
\hypersetup{
    colorlinks,
    citecolor=darkblue,
    filecolor=black,
    linkcolor=magenta,
    urlcolor=black
}

\usepackage[numbers]{natbib}

\newtheorem{definition}{Definition}
\newtheorem{lemma}{Lemma}
\newtheorem{theorem}{Theorem}
\newtheorem{corollary}{Corollary}

\newtheorem{observation}{Observation}

\newcommand{\OPT}{\texttt{OPT}\xspace}
\newcommand{\ALG}{\texttt{ALG}\xspace}

\begin{document}



\title{Multiplicative Bidding in Online Advertising}
\author{MohammadHossein Bateni
\thanks{Google Research, New York. \{bateni, jonfeld, mirrokni\}@google.com
}\and
Jon Feldman\footnotemark[1]\and
Vahab Mirrokni\footnotemark[1]\and
Sam Chiu-wai Wong
\thanks{University of California at Berkeley. samcwong@berkeley.edu}
}



\maketitle

\begin{abstract}
In this paper, we initiate the study of the multiplicative bidding
language adopted by major Internet search companies.  In
multiplicative bidding, the effective bid on a particular search
auction is the product of a base bid and bid adjustments that are
dependent on features of the search (for example, the geographic
location of the user, or the platform on which the search is
conducted).  We consider the task faced by the advertiser when setting
these bid adjustments, and establish a foundational optimization
problem that captures the core difficulty of bidding under this
language.  We give matching algorithmic and approximation hardness
results for this problem; these results are against an
information-theoretic bound, and thus have implications on the power
of the multiplicative bidding language itself.  Inspired by empirical
studies of search engine price data, we then codify the relevant
restrictions of the problem, and give further algorithmic and hardness
results.  Our main technical contribution is an $O(\log
n)$-approximation for the case of multiplicative prices and monotone
values.  We also provide empirical validations of our problem
restrictions, and test our algorithms on real data against natural
benchmarks. Our experiments show that they perform favorably compared
with the baseline.
\end{abstract}

\section{Introduction} \label{sec:intro}



Real-time ad auctions play a vital role in monetizing the Internet.
In a real-time ad auction, bids are entered by the advertiser beforehand,
and the auction is conducted at the time of a pageview or search.
Each individual search query or pageview has a possibly unique set of
features (e.g., geographic location, time of day, device, etc.) that can have a significant effect on the value of the
ad to the bidder, as well as the market price of the ad placement.

Recently, some of the major search engines have started to allow an
advertiser to set {\em bid adjustments} on their ad campaign in order
to account for differences in valuation that are a function of these
types of features~\cite{bid-adjustments-adwords,bid-adjustments-bing}.
Indeed the transition to this mode of bidding has been characterized as one of the most important recent changes to AdWords~\cite{bid-adjustment-media}.
For example, one could set a bid adjustment of $1.1$ for search
queries originating in California, an adjustment of $0.9$ for queries
submitted from 2-3pm, and another adjustment of $1.2$ for mobile
devices.  Then, for a base bid of $\$1$, your final bid on a
California mobile query between 2-3pm would be $\$1 \times 1.1 \times
.9 \times 1.2 = \$1.188$.

Bid adjustments allow an advertiser to express relative valuation
across a supported feature type (for example,
geographic location), but do not allow for specifying valuation on
arbitrary combinations of features.  For example, if an advertiser
found that mobile searches were 30\% more valuable than desktop
searches in New York, but only 15\% more valuable in California, then
this would not be expressible in the language of bid adjustments.
Such limitations are inevitable, as the space of possible
combinations of these features is prohibitively large. There are, of course, other bidding schemes with succinct bid representations. We have chosen to study multiplicative bidding here because it is the status quo. Investigating the expressive power of other schemes is an interesting future direction.

In this paper, we initiate the theoretical and empirical study of
multiplicative bidding and examine the task faced by advertisers given
the option of setting bid adjustments.  We begin by whittling it down
to a simple, elegant optimization problem on two feature
dimensions. This problem still captures the tension of bidding
``multiplicatively,'' rather than individually on each auction, while
ignoring some of the unrelated idiosyncrasies of search ad auctions.
We then fully explore the complexity of this optimization problem.  We
first show that the problem is $\Omega(\sqrt{n})$-hard to approximate,
and give an algorithm which exactly matches this hardness ratio.
Because these results are in relation to a solution where a bidder can
bid individually on each feature combination, this also implies
an information-theoretic limitation of the multiplicative bidding
language itself.

Motivated by analyses on real search auction data, we then examine
the effect of assuming various conditions (e.g., monotonicity on the values and prices in
different dimensions). As our main technical contribution, we develop an $O(\log n)$
approximation algorithm when prices are multiplicative and values are
monotone in one dimension.  We validate our assumptions
on search auction data, and test our algorithm on this data against natural benchmarks.
Before elaborating on our results and techniques, we present a formal
model of the multiplicative bidding problem, and later (in Section~\ref{sec:results}) 
describe details of our theoretical and empirical results.

\subsection{Model}\label{sec:model}
Most search engines conduct some variant of the {\em generalized
second price (GSP)} auction to sell ad placements on user search
queries.  Since we are studying the advertiser-facing budget optimization
problem, an appropriate model for an individual auction would be the
``landscape'' model from~\citet{FMPS07:budget} (see also Section~\ref{sec:related}). 
In this model there is a set of {\em threshold} bids $b_1, \dots, b_n$ (where $n$ is the
number of positions on the page, a small constant), and bidding in
the interval $[b_i, b_{i+1}]$ gives some number of clicks at a cost of $b_i$
per click.  A special case of this model (when $n = 1$) is a
take-it-or-leave-it click at a fixed price $p$.  We will
assume this special case in the present paper for simplicity, since the task
of multiplicative bidding is still sufficiently sophisticated in this
case.  Extending to multiple click supply, multiple ad positions, or multiple queries with
different market prices is an interesting direction for future work.

The bidding dimensions supported by the major search engines include
time of day, geographic location, platform (e.g., mobile device vs.\ desktop) 
and keyword targeting.  Thus it is reasonable to assume that the
number of dimensions is small, but the number of different values in
each dimension is possibly large (for example, AdWords allows bid
multipliers at 15-minute intervals in a week, and at
postal-code geographic level).  A reasonable place to start is the
2-dimensional problem, where each dimension has a large number of
possible values; for the purposes of technical focus, we consider only
this case in the paper, and leave it open to extend our results to
multiple dimensions.

Given these modeling considerations, we now propose the first model
for the multiplicative bidding problem. We feel that this simple model
retains the salient feature of the problem---namely, the tension of
bidding multiplicatively---and is a solid foundation on which to
inspire future work in this area.

\medskip

{\bf \noindent The Multiplicative Bidding Problem.}  Suppose there are two {\em
bid adjustment dimensions} with $m$ and $n$ possible settings,
respectively.  For each \emph{entry} $(i,j)\in [m]\times [n]$, an
advertiser is given a price $p_{ij}>0$ and value $v_{ij}\geq 0$.  He
is required to specify a bid multiplier $r_i$ for each row $i\in [m]$
and $c_j$ for each column $j\in [n]$. The effective bid for cell
$(i,j)$ is then $r_i\cdot c_j$.

A cell $(i,j)$ is said to be \emph{captured} 
if its effective bid is at least the price, i.e., $r_i\cdot c_j\geq p_{ij}$. 
The advertiser also has a budget $B>0$ and is subject to 
the budget constraint $$\sum_{(i,j):r_i c_j\geq p_{ij}}p_{ij}\leq B.$$ 
His objective is to maximize the total value gained 
$$\sum_{(i,j):r_i c_j\geq p_{ij}}v_{ij}.$$

{\bf \noindent Relation to knapsack.}
Multiplicative bidding can be viewed as a restricted version of the classical knapsack problem. 
Indeed, if we were free to bid any amount on each individual cell, 
we would be able to capture any desired subset of the cells, 
where each cell is simply an item with a price and value.  We will refer to this solution as the {\em individual bidding optimum}, or simply, $\OPT$.

With multiplicative bidding, as we shall see, not all subsets of 
the cells can be captured. Consequently, we are optimizing over 
a smaller space, and the best solution available can be no better 
than the individual bidding optimum.

{\noindent \bf Approximation benchmarks.}
One of our objectives is to quantify how much efficiency is potentially lost by 
restricting an advertiser to multiplicative bidding (compared to a real-time bid, as is common in Ad Exchanges, for example). In light of this, 
the most natural benchmark would be the optimal individual bidding solution $\OPT$.
We abuse notation by using $\OPT$ to denote both the set of cells in the optimum as well as their total value.

To simplify our presentation, {we will assume that $\OPT$ 
simply chooses the cells with the best $v_{ij}/p_{ij}$ ratio 
while not exhausting the budget $B$}.
This assumption is reasonable in two ways. 
First of all, the reader may have already noticed that the above 
is the well-known $2$-approximation for knapsack and thus we lose 
a factor of at most $2$ by adopting such a solution.
Secondly, in the context of Internet advertising, 
very often each $p_{ij}$ and $v_{ij}$ is small compared to the 
overall spend and value derived, in which case our solution 
is in fact $(1-\epsilon)$-approximate. More specifically, we assume 
that $p_{ij}/B < \epsilon$ for all $(i,j)$. The reader may readily 
verify that this is indeed an $(1-\epsilon)$-approximation. We stress 
again that almost all our results remain valid without this assumption 
since $\OPT$ is a 2-approximation to the true optimum. This assumption 
is merely introduced in order to simplify our exposition and avoid being overly verbose.

Another more benign benchmark would be the multiplicative bidding optimum, which is useful for characterizing the computational hardness of finding a good solution. 
This is different from the former benchmark which carries the flavor of 
``information theoretical'' lower bounds.  Somewhat surprisingly, 
at least in the general case, the optimal approximation ratio is 
essentially the same with respect to the individual bidding optimum (Lemma~\ref{lem:general}) 
and the multiplicative bidding optimum (Lemma~\ref{lem:nphard}).

\subsection{Our contributions and techniques}\label{sec:results}
The most important goal of our work is to establish a
foundational multiplicative bidding problem on which to build algorithmic insight.
Given the prevalence of this new
bidding language, this represents an urgent call to investigate these bidding schemes in 
different scenarios and to design better bid optimization algorithms for them. 

To this end, we first start with the plain formulation of the problem as stated in Section~\ref{sec:model}, 
and we are able to fully characterize its approximability to be $\Theta(\sqrt{n})$.  Our algorithm is greedy in nature and uses the intuition provided by the hardness result.
In order to better model the prices and values that can arise from practice, we 
then consider a number of monotonicity conditions. For instance, we say that the 
values are monotone along the rows if they can be permuted so that $v_{ij}$ increases 
as $i$ increases.  Our results are summarized in Table~\ref{table: results}, which gives the 
complexity of multiplicative bidding in different cases against the \emph{individual bidding} optimum.

Unfortunately, the lower bound does not really improve even if the values and prices 
are monotone for both rows and columns. Nevertheless, we find that the problem becomes 
tractable given monotone value-over-price ratios (along either row or column). This 
prompts us to consider a subclass of solutions, \emph{staircases}, that are always feasible.

Building upon this staircase notion, we obtain the more optimistic approximation ratio 
of $O(\log n)$ assuming that prices are multiplicative, i.e., $p_{ij}=p_iq_j$ and values 
are monotone along one dimension. These assumptions are justified by empirical data validation
(see Section~\ref{sec:validation}). At a high level, our algorithm attempts to extract a large subset of the optimum and patch it into a staircase. However, care must be taken to avoid overspending. Indeed, the factor $O(\log n)$ is a compromise between the budget constraint and staircase feasibility.

To apply our algorithms in practice, we must deal with the fact that our monotonicity assumptions hold only in an approximate sense.  We address this in Section~\ref{sec:exp} by providing more robust adaptations of two of our algorithms; these adaptations allow the algorithms to work in a general setting, but still take advantage of the near-monotonicity of the data.  We evaluate these algorithms on real search auction data, and show that both have a significant gain over a benchmark inspired by~\citet{FMPS07:budget}.  

\begin{table}
       {%
\begin{center}
\caption{Lower bounds and algorithmic results on the approximation ratio in different cases.}{
       \begin{tabular}{|c|c|c|c|c|c|}
                \hline
                & General 
                & \vtop{\hbox{\strut Monotone value-}\hbox{\strut -over-price}}
                & \vtop{\hbox{\strut Monotone prices}\hbox{\strut and values}}
                & \vtop{\hbox{\strut Multiplicative}\hbox{\strut prices}}
                & \vtop{\hbox{\strut Multiplicative prices}\hbox{\strut and monotone values}}
                \\
                \hline
                Hardness 
                & $\Omega(\sqrt{n})$ \tablefootnote{We have also shown that it is $\Omega(n^{1/2-\epsilon})$-hard to approximate against the less stringent \emph{multiplicative bidding} optimum (Lemma~\ref{lem:nphard}).}
                & 1 
                & $\Omega(n^{1/2-\epsilon})$
                & $\Omega(\sqrt{n})$
                & 1 
                \\

                & Lemma~\ref{lem:general}
                & 
                & Lemma~\ref{lem:monotone2}
                & Lemma~\ref{lem:general}
                & 
                \\
                \hline
                Algorithm 
                & $O(\sqrt{n})$
                & 1
                & $O(\sqrt{n})$
                & $O(\sqrt{n})$
                & $O(\log n)$
                \\

                & Theorem~\ref{thm:general}
                & Corollary~\ref{cor:ratio}
                & Theorem~\ref{thm:general}
                & Theorem~\ref{thm:general}
                & Theorem~\ref{thm:log}
                \\
                \hline
       \end{tabular}}

\label{table: results}
\end{center}}
\end{table}

\subsection{Related work}\label{sec:related}
This work is most related to the paper of \citet*{FMPS07:budget}
in which 
the authors propose uniform bidding as a means for bid optimization in the presence
of budget constraints in sponsored-search ad auctions.
There are several differences between this paper and the previous line of work
on uniform bidding.
Most notably in the multi-dimensional settings, we cannot apply the results of
\citet{FMPS07:budget} and \citet{mps}. 
In fact, as we will observe, our problem in general is 
inapproximable even for a simple setting.

Previous work
on uniform bidding strategies assume that the number of impressions or 
clicks that the advertiser
gets varies as a function of the bid. Here, on the other hand, we assume
that bidding results in winning or not winning the impression. While our
hardness results directly apply to such more general settings, our approximation algorithm results need an extra step to generalize
to this setting. We leave this as an interesting future research direction.

As a central issue in online advertising,  optimizing under budget constraints
has been studied extensively both from publishers' (or search engines') point 
of view~\cite{MSVV,devanur-hayes,GMP,CCCDW13,KMS13}, and from advertisers' point 
of view~\cite{etesami,FMPS07:budget,David,Deep,mps,EMMMN,AMM}. 
More closely relevant to this paper, the bid optimization with budget constraints 
has also been studied from advertisers' perspective: This has been considered either 
in a repeated auction setting~\cite{etesami}, or in the context of broad-match ad 
auctions~\cite{EMMMN}, or the case of long-term carryover effects~\cite{AMM}.

\section{Hardness results}
We present lower bounds of $\Omega(\sqrt{n})$ and $\Omega(n^{1/2-\epsilon})$ 
for the multiplicative bidding problem in various natural scenarios. Besides 
implying that the approximation algorithm in the next section is asymptotically 
optimal, they also show that the requirements enforced on values and prices in 
our $O(\log m)$-approximation cannot be easily loosened.

\begin{lemma} \label{lem:general}
There exists an instance such that the gap between multiplicative bidding and 
individual bidding is $\Omega(\sqrt{n})$, even when the prices are all equal.
\end{lemma}
\begin{proof}
Consider the following bad instance where $m=n$:
\begin{itemize}
\item prices: $p_{ij}=1$
\item values: $v_{ii}=1$, $v_{ij}=0$
\item budget: $B=n$
\end{itemize}
\vspace{-1.5cm}
$$(v_{ij})=\left(\begin{array}{ccc}
1 &  & 0\\
 & \ddots\\
0 &  & 1
\end{array}\right)$$

Hence, we have $\OPT = n$ by picking all diagonal cells. 
We make a crucial observation that realizes the tension to bid multiplicatively. This will be recurrent in this paper:

\begin{observation}
If our bid multipliers capture both $(i,i)$ and $(j,j)$ entries, 
then at least one of $(i,j)$ and $(j,i)$ is also captured.
\end{observation}

The reason for this is simple. Capturing both $(i,i)$ and $(j,j)$ implies 
that $r_i\cdot c_i\geq p_{ii}=1$ and $r_j\cdot c_j\geq p_{jj}=1$. Thus, 
we must have $r_i\cdot c_j\geq p_{ij}=1$ or $r_j\cdot c_i\geq p_{ji}=1$. 
In fact, all we need is $p_{ii}p_{jj}=p_{ij}p_{ji}$.

Now we return to the main proof. Suppose that we collect $k$ diagonal entries. 
Then we must also collect at least $k\choose 2$ off-diagonal entries by the 
observation. Therefore, 
$$B=n\geq {k\choose 2} \implies \ALG=k=O(\sqrt{n})=O(\OPT/\sqrt{n}).\qed$$
\end{proof}

In light of the last lemma, one might hope that some less pessimistic approximation ratio is attainable under certain conditions. We rule out one such possibility.

Prices and values are \emph{monotone} in both dimensions if the rows and columns can be rearranged so that for any $i\geq i',j\geq j'$, we have $v_{ij}\geq v_{i'j'}$ and $p_{ij}\geq p_{i'j'}$.

\begin{lemma} \label{lem:monotone}
There exists an instance such that the gap between multiplicative bidding and individual bidding is $\Omega(n^{1/3})$, even when the prices and values are monotone.
\end{lemma}
\begin{proof}
The proof is a more elaborate version of Lemma \ref{lem:general}. Consider the following bad instance where $m=n$:
\begin{itemize}
\item $p_{ij}=n^{-1/3},v_{ij}=0$ for $i+j<n+1$ (above antidiagonal)
\item $p_{ij}=1,v_{ij}=1$ for $i+j=n+1$ (on antidiagonal)
\item $p_{ij}=n^{1/3},v_{ij}=1$ for $i+j>n+1$ (below antidiagonal)
\item $B=n$
\end{itemize}
$$(p_{ij})=\left(\begin{array}{ccc}
n^{-1/3} &  & 1\\
 & \iddots\\
1 &  & n^{1/3}
\end{array}\right),\;\;\;(v_{ij})=\left(\begin{array}{ccc}
0 &  & 1\\
 & \iddots\\
1 &  & 1
\end{array}\right)$$

Again, we have $\OPT = n$ by picking all antidiagonal entries. Our goal is to show that no algorithm achieves a total value of $\omega(n^{2/3})$.

Let $k$ be the number of antidiagonal entries captured. Further let $k_1$ and $k_2$ be the numbers of entries captured above and below the antidiagonal respectively. As in Lemma \ref{lem:general}, we have $k_1+k_2\geq {k\choose 2}$.

On the other hand, the budget constraint dictates that $$k+ k_1 n^{-1/3} + k_2 n^{1/3}\leq n,$$

which gives $k_1\leq n^{4/3},k_2\leq n^{2/3}$ and in turn $k=O(n^{2/3})$. Our claim is immediate since $\ALG=k+k_2=O(n^{2/3})$.
\end{proof}

We next generalize the construction above to establish an 
$\Omega(n^{\frac{c-1}{2c-1}})$ gap for an arbitrary positive integer $c$. 
Thus the case here corresponds to $c=2$. By taking $c\to \infty$, 
we conclude a hardness of $\Omega(n^{1/2-\epsilon})$.

\begin{lemma} \label{lem:monotone2}\label{LEM:MONOTONE2}
There is an instance such that the gap between multiplicative bidding and individual bidding is $\Omega(n^{1/2-\epsilon})$, even when the prices and values are monotone.
\end{lemma}
\begin{proof}

The bad instance used here extends the one in the proof of Lemma~\ref{lem:monotone}. Consider the following where $m=n$ and $c\geq 2$ is a fixed integer:
\begin{itemize}
\item $p_{ij}=n^{\frac{1}{1-2c}},v_{ij}=0$ for $i+j<n+1$ (above antidiagonal)
\item $p_{ij}=1,v_{ij}=1$ for $i+j=n+1$ (on antidiagonal)
\item $p_{ij}=n^{\frac{c-1}{2c-1}},v_{ij}=1$ for $i+j>n+1$ (below antidiagonal)
\item $B=n$
\end{itemize}

$$(p_{ij})=\left(\begin{array}{ccc}
n^{\frac{1}{1-2c}} &  & 1\\
 & \iddots\\
1 &  & n^{\frac{c-1}{2c-1}}
\end{array}\right),\;\;\;(v_{ij})=\left(\begin{array}{ccc}
0 &  & 1\\
 & \iddots\\
1 &  & 1
\end{array}\right)$$

Thus, $\OPT = n$ by picking all antidiagonal entries. We will show that $\ALG=O(n^{\frac{c}{2c-1}})$. Again, let $k$, $k_1$ and $k_2$ be the numbers of entries captured on, above and below the antidiagonal respectively. We begin by re-establishing the inequality $k_1+k_2\geq {k\choose 2}$.

Consider $c$ antidiagonal entries $(i_1, j_1), \cdots, (i_c, j_c)$, where $i_1 < \cdots < i_c$. We claim that if all of these $c$ entries are captured, then so is at least one of $(i_1,j_2),(i_2,j_3),\cdots,(i_{c-1},j_c),(i_c,j_1)$.

This is clear by noticing that the product of the prices for the first group of entries equals 1, which is the same as the second group's 
$\left(n^{\frac{1}{1-2c}}\right)^{c-1} \cdot n^{\frac{c-1}{2c-1}}=1$. An illustration with $c=3$ is provided below, where (*) and (\#) denote the antidiagonal entries chosen and the corresponding off-antidiagonal entries.

$$\left(\begin{array}{cccccc}
\\
 &  & \# &  & *\\
\\
 & \# & *\\
 & * &  &  & \#\\
\\
\end{array}\right)$$

Now for each group of $c$ antidiagonal entries captured, we must have captured some corresponding off-antidiagonal entry. There are $k\choose c$ such groups. Since each off-antidiagonal entry identify two members of such a group, it corresponds to at most $k\choose {c-2}$ groups. In other words, there are $$k_1+k_2\geq {k\choose c}\Big/ {k\choose {c-2}}=\Theta(k^2)$$ off-antidiagonal entries.

The rest of the proof is similar to before. 
The budget constraint gives us $$k + k_1 n^{\frac{1}{1-2c}}+k_2 n^{\frac{c-1}{2c-1}}\leq n.$$ Combining both inequalities yields $\ALG=O(n^{\frac{c}{2c-1}})$. Finally, our proof finishes by letting $c\to\infty$.
\end{proof}

Finally, we show that the approximation ratio is still dismal even when compared against the less stringent multiplicative bidding optimum.

\begin{lemma} \label{lem:nphard}
Unless $\mathbf{NP}\subseteq \mathbf{BPP}$, there is no randomized polynomial-time 
algorithm that finds a solution of value within a factor of $O(n^{\frac{1-\epsilon}{2}})$ 
of the optimal multiplicative bidding solution.
\end{lemma}
\begin{proof}
This is based on the well-known hardness of $n^{1-\epsilon}$ for maximum independent set~\cite{hastad1999clique}. Given a graph $G=(V,E)$ with the maximum independent set of size $|S|\geq n^{1-\epsilon/4}$, we consider the $n$ multiplicative bidding instances below, where $m=n=|V|$:

\begin{itemize}
\item $p_{ii}=1,v_{ii}=1$
\item $p_{ij}=1,v_{ij}=0$ for $(i,j)\in E$
\item $p_{ij}=0,v_{ij}=0$ for $(i,j)\notin E$
\item $B=|S|$
\end{itemize}

In other words, the values and prices respectively form the identity matrix $I_n$ and $I_n + A(G)$ where $A(G)$ is the adjacency matrix of $G$. It should be clear that the multiplicative bidding optimum simply captures all $S\times S$ entries and achieves a total value of $|S|$.

Suppose that some polynomial time algorithm finds a solution of value at least $|S|/n^{1-\epsilon}$. Let $S'\subseteq [n]$ be the diagonal entries captured. Then $$|S'|\geq \frac{|S|}{n^{(1-\epsilon)/2}.}$$

By the same argument in the proof of Lemma \ref{lem:general}, the algorithm also captures at least one of $(i,j)$ and $(j,i)$ for every $i,j\in S'$. Therefore the number of edges within $S'$ cannot exceed the budget $|S|$, i.e., $$E(S')\leq |S|.$$

Now we exhibit an independent subset $S''\subseteq S'$ of size $n^{\epsilon/2}$. This would lead to an $n^{1-O(\epsilon)}$-approximation for maximum independent set.

Consider a random subset $S''\subseteq S'$ of size $n^{\epsilon/9}$. We  bound the expected number of edges within $S''$. Note that each edge within $S'$ falls inside $S''$ with probability less than $n^{\epsilon/4}/|S'|^2$. Thus the expected number of edges is less than $$\frac{n^{\epsilon/4}}{|S'|^2}\cdot E(S')\leq n^{\epsilon/4}\cdot \frac{n^{1-\epsilon}}{|S|^2}\cdot |S|\leq n^{-\epsilon/2},$$ where we used $|S|\geq n^{1-\epsilon/4}$. This shows that $S''$ is independent with high probability.
\end{proof}

\section{Tight $O(\lowercase{\sqrt{n}})$-approximation algorithm for the general case}
In this section, we present an $O(\sqrt{n})$-approximation for the 
multiplicative bidding problem. This matches the lower bound in 
Lemmas~\ref{lem:general} and~\ref{lem:nphard}. Our algorithm will 
greedily construct at most $O(\sqrt{n})$ disjoint feasible solutions 
whose union captures all the cells in the individual bidding optimum $\OPT$. 
The approximation promise follows immediately by picking one among these.

We first give an overview of the class of solutions found by the algorithm 
before describing it formally. Each of the solutions will focus on certain 
\emph{active} columns. This can be done by bidding 0 on the other columns.

On the other hand, the bid on each active column is just 1, and the bid on 
row $i$ equals the maximum price $p_{ij}$ among the entries $(i,j)\in \OPT$ 
in an active column $j$, or simply 0 if there is no such entry.

Observe that this kind of bidding allows us to capture all the $\OPT$ entries 
in the active columns. Our algorithm is stated as Algorithm~\ref{alg:general}. 
Notice that the candidate solutions $V$ in the while loop are constructed in a 
greedy manner.

\begin{algorithm}
\SetAlgoLined
\label{alg:general}
\caption{$O(\sqrt{n})$ approximation}
$U\longleftarrow [n]$\;
$candidates\longleftarrow\emptyset$\;
\While{$|U|>2\sqrt{n}$}
{
Let $V\subseteq U$ be a maximal set of at most $\sqrt{n}$ columns such that $$\sum_{i=1}^m \max_{(i,j)\in \OPT,j\in V} p_{ij}\leq \frac{B}{|V|},$$ where the $\max$ is 0 if there is no cell from $\OPT$ in row $i$ and columns in $V$\;

Add $V$ to $candidates$\;

Remove $V$ from $U$\;
}
For each remaining $j\in U$, add $V=\{ j\}$ to $candidates$\;
Output the best $V$ from $candidates$
\end{algorithm}

For the best candidate set $V$ in the algorithm, we bid as follows:
\begin{itemize}
\item For $j\notin V$, $c_j=0$.
\item For $j\in V$, $c_j=1$.
\item For $i\in [m]$, $\displaystyle r_i=\max_{(i,j)\in \OPT,j\in V} p_{ij}$, or 0 if $\forall j\in V$, $(i,j)\notin \OPT$.
\end{itemize}

As mentioned before, this bidding scheme captures all the $\OPT$ cells from 
columns in $V$ by design. We first show that such a solution is indeed feasible.

\begin{lemma}
For each candidate $V$, the total spend is at most $B$.
\end{lemma}
\begin{proof}
If $V$ is chosen in the while loop, then for row $i$ we spend at most 
$\displaystyle |V|r_i=|V|\max_{(i,j)\in \OPT,j\in V} p_{ij}$. Summing 
over all rows, we get 
$$\sum_{i=1}^m |V|\max_{(i,j)\in \OPT,j\in V} p_{ij}\leq |V|\cdot\frac{B}{|V|}=B.$$
On the other hand, it is clear that the solution for $V$ chosen after 
the while loop is a subset of $\OPT$ and hence costs at most $B$.
\end{proof}

Next we establish the approximation guarantee. The intuition behind it
is that when $V$ is not large, it consumes a relatively large amount of the budget and hence there cannot be too many $|V|<\sqrt{n}$.

\begin{lemma} \label{lem:bigcost}
If $|V|<\sqrt{n}$ and $V$ is picked before the last candidate in the while loop of the algorithm, we have 
\[
\sum_{i=1}^m \max_{(i,j)\in \OPT,j\in V} p_{ij}\geq \frac{B}{4|V|}.\tag{*}
\]
 In particular, the $\OPT$ entries from columns in $V$ cost at least $B/4|V|$.
\end{lemma}
\begin{proof}
By the maximality of $V$, the fact that $V$ has not reached the size $\sqrt{n}$ implies that no column can be added to $V$.

Note that inserting a new column $j$ to $V$ increases the L.H.S. of (*) by at most $\sum_{i:(i,j)\in \OPT} p_{ij}$, i.e., the amount \OPT spends on $j$.

Since $V$ is not the last, there must be more than $2\sqrt{n}$ available columns in $U$, one of which costs no more than $B/2\sqrt{n}\leq B/2(|V|+1)$. By the maximality of $V$, this column cannot be added to $V$ and we must then have $$\frac{B}{2(|V|+1)}+\sum_{i=1}^m \max_{(i,j)\in \OPT,j\in V} p_{ij}> \frac{B}{|V|+1},$$ from which our result follows.
\end{proof}

\begin{theorem}\label{thm:general}
The algorithm above achieves an approximation ratio of $O(\sqrt{n})$.
\end{theorem}
\begin{proof}
It suffices to show that there are $O(\sqrt{n})$ candidates $V$ as 
they collectively capture all of $\OPT$. Since there can be at most 
$\sqrt{n}$ of them of size $\sqrt{n}$, and at most $2\sqrt{n}$ 
candidates $V$ after the while loop, it suffices to bound the number 
of candidates $V$ with size smaller than $\sqrt{n}$ in the loop.

Let $a_1, \cdots, a_k$ be their sizes. Then trivially $a_1+ \cdots+ a_k\leq n$. 
By Lemma~\ref{lem:bigcost}, their \OPT entries cost at least 
$B/4a_1+ \cdots+ B/4a_k\leq B$, hence $1/a_1+ \cdots+1/a_k\leq 4$. 
Now by the Cauchy-Schwarz inequality, 
$$4n\geq (a_1+\cdots+a_k)\left(\frac{1}{a_1}+\cdots+\frac{1}{a_k}\right)\geq k^2.$$

This gives $k\leq 2\sqrt{n}$, as desired.
\end{proof}

Finally, we remark that a $O(\sqrt m)$-approximation can be obtained 
by swapping the roles of rows and columns. Combining both gives a 
$O(\min\{\sqrt{m},\sqrt{n}\})$-approximation.

\section{Staircases as the building block}
We introduce an important notion which provides a sufficient condition 
on the feasibility of a solution. As we shall see, it will be helpful 
in deriving approximation algorithms for our problem as this subclass 
of solutions is much easier to work with.

\begin{definition}
A configuration $S\subseteq [m]\times [n]$ is a \emph{staircase} if for any $i,j\in [n]$, 
its subset of cells in column $i$ is a subset or superset of that in column $j$.
\end{definition}

It is clear that replacing columns by rows in the definition makes no difference. The name staircase originates from the fact that the rows and columns can be permuted in such a way that $S$ indeed resembles a staircase (with possibly uneven step sizes).

\begin{lemma} \label{lem:staircase}
Ignoring the budget constraint, a staircase $S$ can always be captured exactly.
\end{lemma}
\begin{proof}
%
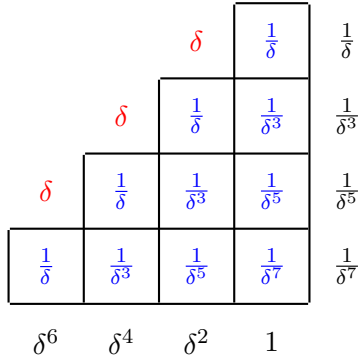
\begin{figure}
\centerline{
\begin{tikzpicture}
\def\scl{1cm}
\tikzstyle{placeholder}=[inner sep=0mm,outer sep=0pt]
\tikzstyle{conn}=[thick]
\tikzstyle{num}=[]
\foreach \x in {0, ..., 4}
  \foreach \y in {0, ..., 4}
    \draw (4 * \scl - \x * \scl, \y * \scl) node[placeholder] (V\x\y) {};
\draw[conn] (V40) -- (V00) -- (V04);
\foreach \x in {1, ..., 4}
{
  \pgfmathtruncatemacro{\z}{5-\x}
  \draw[conn] (V\x0) -- (V\x\z);
  \draw[conn] (V0\x) -- (V\z\x);
  \pgfmathtruncatemacro{\ex}{4*2+1-2*\x}
  \ifnum\ex=1
  \def\ex{}
  \fi
  \draw (V0\x) ++(\scl / 2, -\scl / 2) node[num] {$\frac{1}{\delta^{\ex}}$};
  \pgfmathtruncatemacro{\ex}{2*\x-2}
  \draw (V\x0) ++(\scl / 2, -\scl / 2) node[num] {
    \ifnum\ex=0
    $1$
    \else
    $\delta^{\ex}$
    \fi
  };
  \foreach \y in {1, ..., \z} {
    \pgfmathtruncatemacro{\ex}{4*2+1-2*\y -2*\x+2}
    \ifnum\ex=1
    \def\ex{}
    \fi
    \draw (V\x\y) ++(\scl / 2, -\scl / 2) node[num,blue] {$\frac{1}{\delta^{\ex}}$};
  }
  \ifnum\x<4
  \draw (V\x\z) ++(-\scl/2,-\scl/2) node[num,red] {$\delta$};
  \fi
}
\end{tikzpicture}
}
\caption{A \emph{regular} staircase with four rows and four columns.
The numbers shown at the bottom and to the right are the column and row multipliers respectively.
The numbers shown in blue inside the staircase indicate the effective bids with a lower bound
of $1/\delta$. The numbers shown in red outside the staircase are the effective bids
with an upper bound of $\delta$.
Note that each row or column can have different height or width
if we repeat the associated multiplier.\label{fig:staircase}%
}
\end{figure}

Figure~\ref{fig:staircase} demonstrates our bidding scheme to capture a simple staircase.
Notice that we are bidding $1/\delta^i$ within the staircase and $\delta^i$ outside, 
where $i$ is a positive integer varying across different cells. 
By making $\delta$ sufficiently small, we can capture the staircase exactly.

In general, if $S$ has no entry in any row or column, we set the corresponding bid multiplier to 0 and thus essentially remove it. We then permute the remaining rows and columns such that our staircase takes on a shape as Figure~\ref{fig:staircase}. For the rows, we simply bid $1/\delta,1/\delta^3,\cdots$ successively on each level from top to bottom. In contrast, the bids on the columns are $1,\delta^2,\cdots$ on each level from right to left.
\end{proof}

\subsection{$1$-approximation for the case of monotone $v_{ij}/p_{ij}$}

The staircase idea immediately implies that our problem can be solved (almost) exactly under one natural assumption.

\begin{corollary}\label{cor:ratio}
If the ratios $v_{ij}/p_{ij}$ are monotone in one dimension, then there is a $1$-approximation algorithm.
\end{corollary}
\begin{proof}
This is immediate since $\OPT$, which collects entries with the best $v/p$ ratio 
as long as the budget $B$ has not been exhausted, 
will be a staircase when $v/p$ is monotone in one dimension. 
By Lemma~\ref{lem:staircase}, $\OPT$ can be captured exactly. This can be implemented efficiently by selecting the entries with highest $v/p$ ratios one at a time.
\end{proof}

\section{$O(\lowercase{\log m})$-approximation when prices are multiplicative and values are monotone}
In this section, we make the following two assumptions on the prices and values.
\begin{itemize}
\item Multiplicative prices: there are $p_i,q_j>0$ such that $p_{ij}=p_i\cdot q_j$.
\item Monotone values: the values are monotone along one dimension, say, rows. In other words, the rows can be permuted so that $v_{ij}\geq v_{i'j}$ for $i'>i$.
\end{itemize}

Both of them are necessary in the sense that without either of them, 
no algorithm can have a good performance as demonstrated by the bad instances 
in Lemmas \ref{lem:general} and \ref{lem:monotone2}. 
In the former case, the instance has all prices equal (hence multiplicative) 
and a gap of $\Omega(\sqrt{n})$, whereas in the latter case, the values are 
even monotone in both dimensions but the gap is $\Omega(n^{1/2-\epsilon})$. 
Our assumptions are verified empirically in Section~\ref{sec:exp}.

We now have a nice characterization of the configurations that can be captured.

\begin{lemma}
If prices are multiplicative, a configuration $S\subseteq [m]\times [n]$ can be captured if and only if it is a staircase.
\end{lemma}
\begin{proof}
One direction simply reiterates Lemma \ref{lem:staircase}. For the other direction, let $S$ be a feasible configuration that is not a staircase. Equivalently, there are two columns $j_1$ and $j_2$ for which $S\cap ([m]\times\{j_1\})$ is neither a subset nor a superset of $S\cap ([m]\times\{j_2\})$.

Thus, there are some $i_1$ and $i_2$ such that $(i_1,j_1),(i_2,j_2)\in S$ and $(i_1,j_2),(i_2,j_1)\notin S$. This is a contradiction since the former implies that $(r_{i_1}c_{j_1})(r_{i_2}c_{j_2})\geq (p_{i_1}p_{j_1})(p_{i_2}p_{j_2})$ but the latter gives $(r_{i_1}c_{j_2})(r_{i_2}c_{j_1})< (p_{i_1}p_{j_2})(p_{i_2}p_{j_1})$.
\end{proof}

As a consequence of this lemma, it is sufficient to search for a 
good staircase within our budget. We present an algorithm fulfilling 
this objective step by step, each of which, albeit seemingly unrelated, 
will serve its own purpose. While our derivation establishes an 
approximation of $O(\log m)$, it can be easily turned into an 
$O(\log n)$-approximation by swapping the roles of rows and columns and 
assuming column monotonicity instead of row monotonicity.

\subsection{Algorithm}

Our algorithm consists of three major steps. An overview is also given in Algorithm~\ref{alg:log}.

\subsubsection*{Step 1: Clustering prices}

We round \emph{down} all row price multipliers $p_i$ to powers of 2, and permute the rows such that $p_i$ increases from top to bottom. This results in Figure~\ref{fig:algo:1} where each inner rectangle represents a set of rows with equal $p_i$. We call these inner rectangles \emph{strips}.


Since $p_i$ is now a power of 2, the prices of two cells in the same 
column but consecutive strips differs by a factor of at least 2.

\subsubsection*{Step 2: Finding $\OPT(B/4)$}

As one may have guessed, our algorithm builds upon the optimal solution 
to find a good staircase---with one caveat. Instead of basing our solution 
on the $\OPT$ with budget $B$, we use the $\OPT$ with budget $B/4$, 
denoted by $\OPT(B/4)$. The reason for this choice will become apparent 
in the proof of Lemma~\ref{lem:budget}.

We first permute the rows within each strip such that $v_{ij}$ is increasing as we traverse down.  This is possible because of value monotonicity. Now $\OPT(B/4)$ must consist of width-1 towers that sit on the bottom of our strips, as the prices within the same column in a strip are constant. Figure~\ref{fig:algo:2} gives one possible $\OPT(B/4)$.

In summary, we
\begin{itemize}
\item permute the rows in each strip such that $v_{ij}$'s are increasing from top to bottom; and
\item find the cells in $\OPT(B/4)$, which must emanate from the bottom of a strip.
\end{itemize}


\subsubsection*{Step 3: Constructing a staircase from $\OPT(B/4)$}

The last step of our staircase construction is made up of two main substeps:
we first extract a subset of $\OPT(B/4)$ according to a certain parameter $h\in [m]$ 
and then apply some patching work to transform it into a staircase, denoted $\ALG_h$. 
The final solution $\ALG$ will be the best $\ALG_h$.

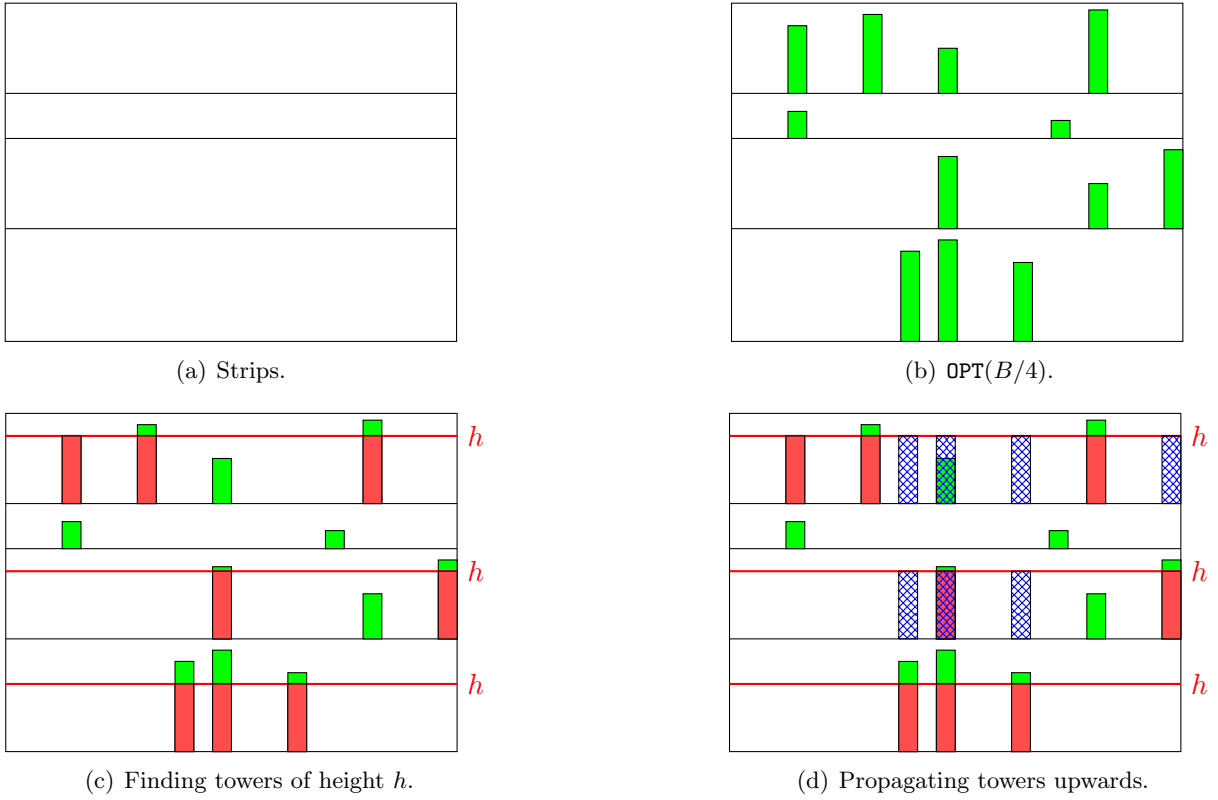
\begin{figure}
{\centering
\def\sclx{5mm}
\def\scly{3mm}
\def\drawboard{
\draw (0, 0) rectangle (12 * \sclx, 15 * \scly);
\foreach \h in {5, 9, 11}
  \draw (0, \h * \scly) -- (12 * \sclx, \h * \scly);
}
\def\drawtower#1#2#3#4{
  \draw[fill=#4] (#1 * \sclx, #2 * \scly) rectangle (#1 * \sclx - \sclx / 2, #2 * \scly + #3 * \scly);
}
\def\drawopt{
\foreach \a/\b/\c in {5/0/4, 6/0/4.5, 8/0/3.5, 
  6/5/3.2, 10/5/2, 12/5/3.5, 
  2/9/1.2, 9/9/0.8,
  2/11/3, 4/11/3.5, 6/11/2, 10/11/3.7}
  \drawtower{\a}{\b}{\c}{green};
}
\def\drawhlines{
\foreach \h in {0, 5, 11} {
    \draw[thick,red] (0, \h * \scly + 3 * \scly) -- (12 * \sclx, \h * \scly + 3 * \scly);
    \draw[thick,red] (12 * \sclx + \sclx/2, \h * \scly + 3 * \scly) node {$h$};
  }
}
\def\drawhtowers{
\foreach \a/\b/\c in {5/0/3, 6/0/3, 8/0/3, 
  6/5/3, 12/5/3, 
  2/11/3, 4/11/3, 10/11/3}
  \drawtower{\a}{\b}{\c}{red!70};
}
\def\drawpropagation{
\foreach \a/\b/\c in {5/5/3, 6/5/3, 8/5/3, 
  5/11/3, 6/11/3, 8/11/3, 12/11/3}
  \drawtower{\a}{\b}{\c}{orange!70,pattern=crosshatch,pattern color=blue};
}
\subfigure[Strips.]{
\label{fig:algo:1}
\begin{tikzpicture}
\drawboard
\end{tikzpicture}
}
\hfill
\subfigure[$\OPT(B/4)$.]{
\label{fig:algo:2}
\begin{tikzpicture}
\drawboard
\drawopt
\draw[thick,white] (12.5 * \sclx, 3 * \scly) node {$ha$};
\end{tikzpicture}
}
\subfigure[Finding towers of height $h$.]{
\label{fig:algo:3}
\begin{tikzpicture}
\drawboard
\drawopt
\drawhlines
\drawhtowers
\end{tikzpicture}
}
\hfill
\subfigure[Propagating towers upwards.]{
\label{fig:algo:4}
\begin{tikzpicture}
\drawboard
\drawopt
\drawhlines
\drawhtowers
\drawpropagation
\end{tikzpicture}
}
}
\caption{The steps in approximating the best multiplicative bidding scheme.
The green ``towers'' in (b) show the cells in $\OPT(B/4)$.
The red lines in (c) illustrate one candidate height $h$.
Note that the second strip is shorter than $h$, and therefore does not
contribute to $\ALG_h$ at all.
The red towers in (c) depict the intermediate solution of height $h$;
this is extended to $\ALG_h$ in (d) via propagating upwards: see the hatched towers.
\label{fig:algo}%
}
\end{figure}

Again, for notational convenience we will use $\ALG_h$ and $\ALG$ to denote both the set of cells in the solution and their total value.
\begin{itemize}
\item \textbf{Disregard strips of height less than $h$}
\item[] No cells in such strips will be chosen.
\item \textbf{Take all height-$h$ towers}
\item[] For the constituent towers of $\OPT(B/4)$ with height at least $h$, we select its $h\times 1$ subtower sitting on the bottom of a strip (Figure~\ref{fig:algo:3}), and insert them into $\ALG_h$.
\item \textbf{Propagate/copy the height-$h$ towers upwards}
\item[] If the $h\times 1$ subtower in strip $i$ and column $j$ is chosen in the previous step, we select the corresponding $h\times 1$ subtower in strip $i'$ and column $j$ for all $i'<i$ provided that strip $i'$ has height at least $h$ (Figure~\ref{fig:algo:4}). They are inserted into $\ALG_h$.
\end{itemize}

Our solution $\ALG_h$ thus consists of all the cells selected above. 
In the figures, these are depicted as red and hatched blue regions. 
Notice that no cells in strip 2 were selected since its height is 
smaller than $h$. A height-$h$ tower can be chosen in both steps.

\textbf{The algorithm outputs the best $\ALG_h$ for $1\leq h\leq m$, i.e., $\ALG=\max_{h\in [m]} \ALG_h$.}

\begin{algorithm}
\SetAlgoLined
\caption{Overview of $O(\log m)$ approximation}
\label{alg:log}
\textbf{Step 1:}

Round down all of the $p_i$'s to the nearest powers of 2\;
Cluster together rows with the same $p_i$\;
Reorder the clusters in increasing $p_i$\;

\textbf{Step 2:}

Reorder the rows within each cluster in increasing values\;
Compute $\OPT(B/4)$, the individual bidding optimum with budget $B/4$\;

\textbf{Step 3:}

\For{h=1,2,\dots,m}
{
$\ALG_h\longleftarrow \emptyset$\;
Insert into $\ALG_h$ all height-$h$ towers of $\OPT(B/4)$\;
Insert into $\ALG_h$ all height-$h$ towers in the strips above the ones in the last line\;
}

Output the best $\ALG_h$ as $\ALG$
\end{algorithm}

\subsection{Analysis}

\begin{lemma} \label{lem:isstaircase}
$\ALG_h$ is a staircase.
\end{lemma}
\begin{proof}
This is almost by design. For each column, the cells in $\ALG_h$ are exactly the height-$h$ towers up to some strip $i$ (excluding those whose height is less than $h$). This is guaranteed by the propagation operation.

Thus the cells of $\ALG_h$ in a column must be a subset or superset of that in another.
\end{proof}

\begin{lemma}\label{lem:budget}
$\ALG_h$ costs no more than $B$, i.e., $\sum_{(i,j)\in \ALG_h} p_{ij}\leq B$.
\end{lemma}
\begin{proof}
First of all, the cells selected in the first substep of step 3 cost at most $B/4$ since they are part of $\OPT(B/4)$. We argue that the propagation operation, which is the second substep, spends also at most $B/4$.

Recall that we have rounded the row price multipliers to powers of 2 in step 1. In particular, the prices between two consecutive strips must then differ by a factor of 2 or more. Therefore a height-$h$ tower, when copied upwards, spends at most $1/2^i$ of its cost on the $i$-th strip above it.

Now summing over all height-$h$ towers from $\OPT(B/4)$, the total cost of the propagation operation does not exceed $B/4\cdot (1/2+1/4+\dots) = B/4$.

Our total cost, which has accounted for all of $\ALG_h$, is bounded by $B/4+B/4=B/2$. Finally, since we have rounded down prices to powers of 2 in step 1, the actual cost of $\ALG_h$ cannot be more than twice of $B/2$.
\end{proof}

\begin{lemma} \label{lem:smallerb}
We have $\OPT(B/4)\geq \OPT\cdot (1/4-\epsilon)$.
\end{lemma}
\begin{proof}
We will use the small cost assumption $p_{ij}/B < \epsilon$ here.\footnote{We note that, as mentioned before, our $O(\log m)$-approximation still holds without it.}

Recall that $\OPT(B')$ collects the cells $(i,j)$ in the order of decreasing $v_{ij}/p_{ij}$ until the budget $B'$ is exhausted. Thus, the extra cells collected in $\OPT\setminus \OPT(B/4)$, which cost $3B/4$, each have a $v_{ij}/p_{ij}$ ratio no better than those in $\OPT(B/4)$.

This implies that $\OPT \leq  \frac{\OPT(B/4)}{1/4-\epsilon}$, since $\OPT(B/4)$ has a cost of more than $B\cdot (1/4 - \epsilon)$ by the small cost assumption.
\end{proof}

\begin{lemma} \label{lem:approx}
We have  $\ALG_h\geq \OPT(B/4)/2\log m$ for some $h$.
\end{lemma}
\begin{proof}
Let $\OPT_h (B/4)$ be the total values of the $\OPT(B/4)$ cells at height $h$ in each strip. We claim that $$\ALG_h\geq h\cdot \OPT_h (B/4).$$

This can be seen as follows. Each cell $(i,j)\in \OPT_h (B/4)$ is the top of some height-$h$ tower in a strip. Notice that the $h-1$ cells below $(i,j)$ have values at least $v_{ij}$ by value monotonicity. Our claim then follows from summing over all $(i,j)\in \OPT_h (B/4)$.

The rest of the proof is standard. Suppose that no choice of $h$ 
gives $\ALG_h\geq \OPT(B/4)/2\log m$. 
Then for all $h$, we have $$h\cdot \OPT_h (B/4)\leq \ALG_h < \frac{\OPT(B/4)}{2\log m},$$
which implies $$\OPT(B/4)=\sum_{h=1}^m \OPT_h (B/4)< \sum_{h=1}^m \frac{\OPT(B/4)}{h\cdot 2\log m}\leq \OPT(B/4).$$

This is a contradiction.
\end{proof}

Combining all the lemmas, we obtain our main result.

\begin{theorem} \label{thm:log}
Our algorithm gives an $O(\log m)$-approximation.
\end{theorem}
\begin{proof}
Lemmas \ref{lem:isstaircase} and \ref{lem:budget} establish the feasibility of $\ALG_h$. The approximation guarantee follows from Lemmas \ref{lem:smallerb} and \ref{lem:approx}.
\end{proof}

We highlight the roles played by different steps of the algorithm. Step 1, which clusters similar price multipliers, ensures that the propagation procedure in Step 3 would result in a convergent geometric sum for the cost. Step 2 computes $\OPT(B/4)$ which is essential to achieving our approximation guarantee as well as the budget constraint. Lastly, the propagation procedure in Step 3 also enforces the staircase feasibility.

Our ratio of $O(\log m)$ might not be the most desirable. However, the algorithm likely outperforms its theoretical guarantee in average case. For example, one reason is that the propagation procedure hatches new towers above existing ones but their values are not accounted for in the analysis as they can be negligible in the worst case. In practice, however, such an extreme pattern of values should be rare.


Finally, we outline an approach that may yield an improved $O\left(\frac{\log m}{\log\log m}\right)$-approximation.

\subsection{Towards $O\left(\frac{\log m}{\log\log m}\right)$-approximation}

We incorporate into our algorithm a heuristic which improves the approximation factor for certain instances. Recall how we are handling the strips. For each of them, we capture all of its height-$h$ sub-towers. Besides guaranteeing good approximation, it ensures that we are not overspending as they are part of $\OPT$.

In retrospect, it is actually possible to achieve an even better performance by capturing some of the remaining entries, while still not violating our budget. We prove the following lemma which subsumes the essence of our algorithm.

\begin{lemma} \label{lem:area}
Let $m$ be a positive integer. Given a piecewise-continuous monotonically decreasing function $f:(0,m]\to\mathbb{R}_+$, let $A=\int_0^m f(t)dt$ be the area under $f$. Then there is some integer $h\in (0,m]$ for which the overlap of the rectangle $[0,h]\times [0,A/h]$ with the region under $f$ is at least $\Theta(A\log\log m/\log m)$.
\end{lemma}
\begin{proof}
Since we can write $f$ as the limit of a sequence of monotonically decreasing continuous functions, we assume for convenience that $f$ is continuous so that $f^{-1}$ is well-defined on $[f(h),f(0))$. Observe the trivial fact $A/h\geq f(h)$.

Consider the sequence defined by $t_1=m$ and whenever $A/t_i\leq f(0_+)$, $$t_{i+1}=f^{-1}(A/t_i).$$ We claim that some $t_i$ satisfies our requirement. Note that the rectangles for these $h=t_i$ cover the entire region under $f$. Let $\alpha = \log\log m/\log m$.

If $t_{i+1}\geq \alpha t_i$ for some $i$, then the intersection between the region under $f$ and the rectangle $[0,t_{i+1}]\times [0,A/t_{i+1}]$ has an area at least $$t_{i+1}\cdot\frac{A}{t_i}\geq A\alpha.$$

If there is no such $i$, then every two successive $t_i$ must be of a factor greater than $\alpha$ apart. Hence our sequence has no more than $\log_\alpha m = \log m/\log\log m$ terms, one of which must give a rectangle that covers a region of area $\Theta(A\log\log m/\log m)$ under $f$.
\end{proof}

Notice that the rectangle $[0,h]\times [0,A/h]$ has area precisely $A$. Informally, it enforces the budget constraint. Furthermore, the fact that the rectangle does not fall completely under $f$ means that we are not just taking towers of height at least $h$, which in turn enhances our performance.

Finally, we remark that the factor $\Theta\left(\frac{\log m}{\log\log m}\right)$ is essentially the best by inspecting the lemma with $f(t)=1/(t+1)$.

\begin{theorem}\label{thm:loglog}
There is an algorithm that delivers a staircase of value at least $$\Theta\left(\frac{\log\log m}{\log m}\right)\cdot B\min_{(i,j)\in \OPT}\frac{v_{ij}}{p_{ij}}.$$
\end{theorem}
\begin{proof}
We adopt the notation of Lemma \ref{lem:area}.

Recall that $\OPT_h(B/4)$ is the set of the cells in $\OPT(B/4)$ at height $h$. Define $f:(0,m]\to\mathbb{R}_+$ by $$f(t)=\sum_{(i,j)\in OPT_h(B/4)} p_{ij}$$ for $h-1 < t \leq h$. Alternately, $f$ is formed by juxtaposing the towers of $\OPT(B/4)$ in the order of their heights. Note that $A=\int_0^m f(t)dt$ is precisely the total price of $\OPT(B/4)$, which is just $B/4$.

By Lemma \ref{lem:area}, there is some $h$ whose intersection with the region under $f$ has area at least $$\Theta\left(A\frac{\log\log m}{\log m}\right)=\Theta\left(B\frac{\log\log m}{\log m}\right).$$

Now by taking the height-$h$\footnote{It is possible that the tower resides in a strip of height less than $h$, in which case we would just take it top-to-bottom.} towers corresponding to the intersection, we spend at most $A=B/4$. On the other hand, the total value of our solution is at least $$\Theta\left(B\cdot \frac{\log\log m}{\log m}\right)\cdot \min_{(i,j)\in \OPT}\frac{v_{ij}}{p_{ij}} = \Theta\left(\frac{\log\log m}{\log m}\right)\cdot B\min_{(i,j)\in \OPT}\frac{v_{ij}}{p_{ij}}.$$

Now the rest of the proof follows in the same manner as Theorem \ref{thm:log}.
\end{proof}

In particular, if the ratios $\frac{v_{ij}}{p_{ij}}$ are close to being uniform in $\OPT$, 
we have an $O\left(\frac{\log m}{\log\log m}\right)$-approximation.

\section{Empirical study}\label{sec:exp}
In this section, we report the results of our experiments. We start by explaining the algorithms implemented, followed by a discussion on their empirical performances. An analysis is also given on the validity of the assumptions made throughout the paper.

\subsection{Algorithms and implementation notes}

We have tested our algorithms on real world data and compared their performances. 
We discuss how they are implemented in the rest of this section.

One of the recurrent themes in our paper is monotonicity. 
In the real world, however, data rarely obey monotonicity perfectly 
due to presence of noise, among other factors. To apply our algorithms, 
we must still somehow obtain a decent ordering of the rows (or columns) 
so that monotonicity approximately holds.

For instance, our monotone value-over-price ratio algorithm from 
Corollary~\ref{cor:ratio} assumes that $v_{ij}/p_{ij}$ increases 
along the columns (or rows). In essence, each of the $n$ columns 
induces one possible (partial\footnote{The permutations are sometimes partial since the values of a cell can be missing due to data sparsity.\label{note:partial}}) permutation of the rows and our job 
is to aggregate these $n$ candidates into one representative 
``consensus permutation''. To this end, we adopt the procedure 
in Algorithm~\ref{alg:order}. This is also used as a subroutine 
of our $O(\log n)$-approximation implementation to rank values. See section~\ref{sec:monoval} for a more in-depth discussion of the method.

\subsubsection{Staircase algorithm: $1$-approximation for Monotone value-over-price ratios}

The discussion above sums up how we permute the rows. 
In the theoretical algorithm, the next step is to simply take $\OPT$, 
which is a staircase, as the solution. But as mentioned before, 
this is not always feasible due to imperfect monotonicity. 
In the implementation, we resort instead to finding a good staircase. 
More specifically, our solution captures a number of consecutive cells 
at the bottom of each column.

Thus now we have $m$ possible choices for each column. 
The resulting optimization problem can be solved by a dynamic program 
that computes the best solutions formed by the first $j$ columns 
for different budget thresholds.

\subsubsection{Tower building algorithm: $O(\log n)$-approximation for multiplicative prices and monotone values}

This is similar to the last implementation. 
We first cluster together rows with similar prices. 
For each cluster we then find a permutation of its rows so that values are roughly monotone. 
At this point we encounter the same issue, 
namely that $\OPT(B/4)$ is not necessarily a staircase in each price cluster. 
Moreover, the factor $1/4$ for budget scaling is a rather arbitrary parameter 
to make the formal analysis go through. Our implementation employs 
a different view of the algorithm via dynamic programming.

For a given column, our algorithm has the freedom to choose the 
height-$h$ towers in the first $i$ price clusters, where $i$ 
ranges from $0$ to the number of clusters. Hence there are at 
most $m$ choices per column. Now the dynamic program simply recursively 
computes the best solution formed by the first $j$ columns 
for different budget thresholds.

\subsubsection{Uniform bidding}
We use an algorithm similar to that proposed by \citet{FMPS07:budget}
as a baseline of how well our algorithms perform.  Feldman et al.\ show that in a
large class of bidding problems\footnote{They define their model based on how bids
translate to costs and clicks in a collection of second-price auctions by looking
at the so-called bid landscapes.}, 
using a single bid on all keywords
guarantees a $1-1/e$ approximation, and provides much better performance
in practice.

In our adaptation of that algorithm, we find the largest single bid $b$ that can be placed on
all cells in the table while respecting the budget constraint.  Clearly this can
be realized by multiplicative bidding.

\subsection{Results and comparison}

We use a dataset of $1000$ randomly selected anonymized advertisers
using Google AdWords system.
Two of the dimensions that these advertisers can provide bid multipliers
for are ``geo'' and ``keyword.''  The types of impressions that the
advertisers are interested in has a wide range as well: from a few up to
hundreds of different geo locations, and from tens to thousands of
keywords.
For our experiments, conversion data acts as a proxy for value,
and historic cost-per-click data is used to derive sample price.

\begin{figure}
\centerline{\includegraphics[trim=0.38cm 0.38cm 0.39cm 1.61cm, clip=true, width=8cm]{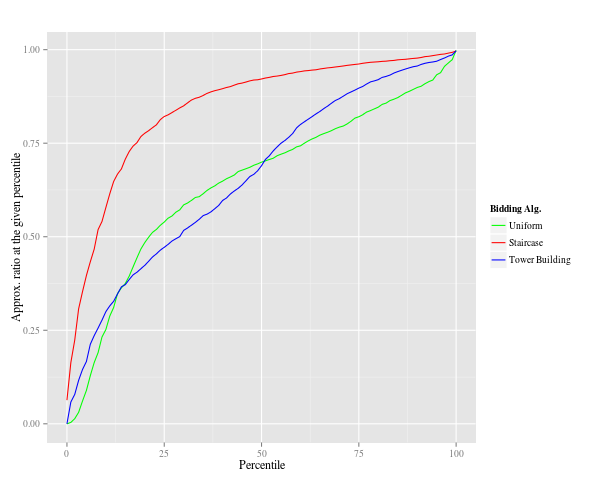}}
\caption{The performance of the three algorithms.
The plot shows what percentage of the individual bidding optimum
each algorithm achieves, and quantifies it based on fraction of
instances.  For instance, at the $50\%$ percentile, in half the instances the performance of the
three algorithms is $69\%, 69\%, 92\%$, respectively.}
\label{fig:perf-histo}
\end{figure}

We first plot the performance of the three different algorithms, 
namely uniform bidding, $O(\log n)$ approximation (also called ``tower building'')
and ratio-based optimization (also called ``staircase'').
The three curves in Figure~\ref{fig:perf-histo} show the percentage of the
individual bidding optimum that each algorithm can obtain.
The uniform bidding approach guarantees $64\%$ of the optimum on average,
while this number is $66\%$ and $85\%$ for $O(\log n)$ approximation and
ratio-based optimization, respectively.
The median performances for the three algorithms are 
$69\%, 69\%, 92\%$ respectively.

\begin{figure}
\centerline{\includegraphics[trim=0.38cm 0.31cm 0.5cm 1.2cm, clip=true, width=7.75cm]{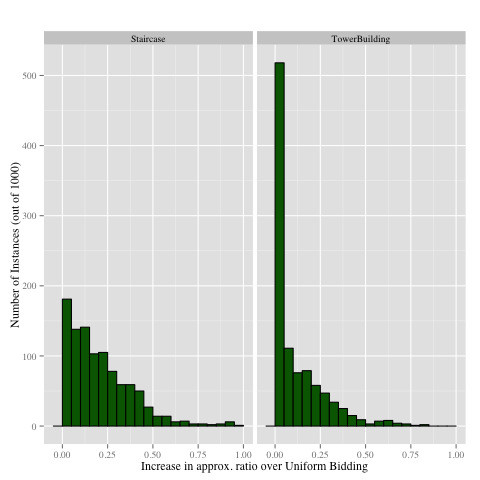}}
\caption{The gain of our algorithms with respect to uniform bidding.
For example, at $0.25$ on the $x$ axis,
we can read the number of instances where either algorithm could improve uniform bidding
by $25\%$.}
\label{fig:gain-histo}
\end{figure}
These numbers suggest that the staircase approach
is much better than the uniform bidding,
whereas the tower building algorithm does not give us any benefits
over the uniform bidding.
The latter conclusion is not entirely true.
Investigating how often the tower building algorithm outperforms
the uniform bidding result, we find that the best of the two algorithms
provides a mean of $75\%$ and a median of $81\%$ for performance.
Figure~\ref{fig:gain-histo} illustrates this with a histogram
of the gain of the two algorithms over the uniform bidding result.
\begin{figure}
\centerline{\includegraphics[trim=0.37cm 0.29cm 0.7cm 1.59cm, clip=true, width=8cm]{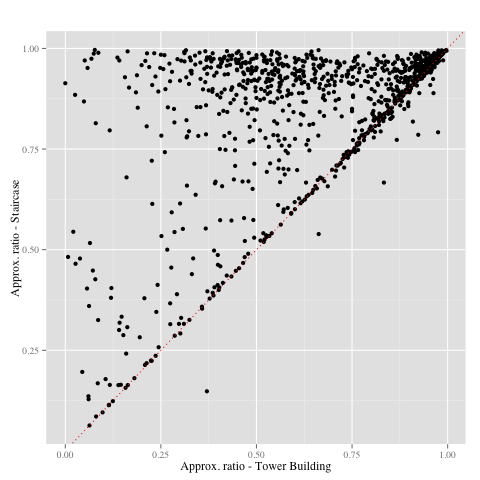}}
\caption{Comparison between our two algorithms: tower building vs. staircase}
\label{fig:compar}
\end{figure}
Finally we compare the staircase and tower building algorithms.
As expected from the above analysis, the former is almost always the better approach.
Though the latter outperforms the former in $10\%$ of the instances
in our dataset, the gains in these cases are nominal.  See Figure~\ref{fig:compar}
for details.

\subsection{Validating our assumptions}\label{sec:validation}

\subsubsection{Are prices multiplicative?}

To validate the assumption that prices are multiplicative, we looked
at click price data from AdWords, aggregating by the country in which
the search query was performed, and the hour of the day.  (We did this
aggregation over all ad clicks over a week of time in November, 2013.)
From this we obtained $p_{ij}$, the average click price for each
combination of country $i$ and hour $j$.  We then looked for two
vectors $r$ and $c$ such that the price $p_{ij}$ was well-approximated
by $r_i c_j$.  To find this vector we ran a linear regression fitting
$\log(p_{ij}) \sim \log(r_1) + \dots + \log(r_n) + \log(c_1) + \dots +
\log(c_m)$, which results in independent coefficients for each country
and each hour of the day.  If this is a good fit, it means
that $\log(p_{ij}) \approx \log(r_i) + \log(c_j)$, which implies
$p_{ij} \approx r_i c_j$.

\begin{figure}
\centerline{\includegraphics[trim=0.37cm 0.30cm 0.89cm 1.52cm, clip=true, width=6.5cm]{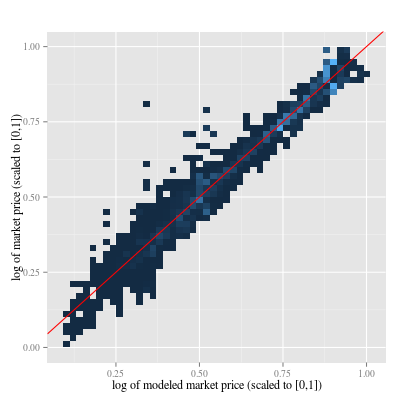}}
\caption{Verifying that prices are multiplicative.  For all $i,j$, the actual market price $p_{ij}$ ($y$-axis) is plotted against a 
predicted market price $r_i c_j$ ($x$-axis), where $r_i$ and $c_j$ are computed using linear regression.  Prices are shown in log scale, under a linear transformation to $[0,1]$.  \label{fig:multcheck}}
\end{figure}

The regression indeed was a good fit, with virtually every country and
hour as significant predictors, and an $R^2$ value of $0.94$.  The
density plot in Figure~\ref{fig:multcheck} shows the actual prices vs.\ the 
 prices predicted by the regression model, i.e., it plots $p_{ij}$ vs. $r_i
c_j$ (note the scale has been changed to $[0,1]$ for privacy purposes).

\subsubsection{Are values and ratios monotone?}\label{sec:monoval}
For $1\leq j\leq n$, let $S_j\subseteq [m]$ be the set of entries in column $j$, 
and $\pi_j$ be the permutation of $S_j$ induced by increasing order of the values (or ratios) of $S_j$.
In other words, we have $n$ partial\footref{note:partial} permutations of $[m]$ 
which we wish to aggregate into one ``consensus'' permutation $\pi$.

Our heuristic in Algorithm~\ref{alg:order} was inspired by the algorithm for rank aggregation
in \citet{ailon2008aggregating}, where the input orderings $\pi_j$ are complete 
rather than partial.

\begin{algorithm}
\SetAlgoLined
\caption{Heuristic for computing consensus permutation}
\label{alg:order}
\KwIn{partial permutations $\pi_1,\dots,\pi_n$}
\KwOut{total ordering $\pi$}

Construct a digraph $D$ over the vertex set $[m]$:

\For{$i,i'\in [m],i\neq i'$}{
\If{no $\pi_j$ contains both $i,i'$}{
continue\;
}
\uIf{more than half the $\pi_j$'s containing $i,i'$ have $i <_{\pi_j} i'$}{
insert edge $(i,i')$ to $D$\;
}
\Else{
insert edge $(i',i)$ to $D$\;
}
}

Maintain a partial order $\pi$ initialized to be empty:

\For{$i=1,2,\dots,m$}{
Let $s_i \in [m]$ be a random vertex of $D$ not processed yet.

\For{each parent $p$ of $s_i$}{
make $p <_\pi s_i$ in $\pi$ if $p,s_i$ are currently incomparable\;
}
\For{each child $c$ of $s_i$}{
make $c >_\pi s_i$ in $\pi$ if $c,s_i$ are currently incomparable\;
}
}

\While{$\pi$ is not yet a total order}{
find two incomparable $i,i'\in [m]$ and make $i <_\pi i'$ in $\pi$\;
}
\end{algorithm}

In words, the heuristic first constructs a digraph encoding the dominance relationship 
between every two $i,i'\in [m]$ by a majority vote. Then a random vertex is selected 
one at a time and used to extend our partial order $\pi$. At the end, the remaining 
incomparable pairs are ordered arbitrarily.

We evaluate how good $\pi$ is with respect to $\pi_1,\dots,\pi_n$ 
by the fraction of agreements:
$$\text{quality}(\pi;\pi_1,\dots,\pi_n)=\frac{\sum_{j=1}^n \#\{(i,i') \in S_j \times S_j| i <_{\pi_j} i', i <_\pi i'\}}{\sum_{j=1}^n {|S_j|\choose 2}}.$$

Thus we have $\text{quality}(\pi;\pi_1,\dots,\pi_n)=1$ if our total order $\pi$ is perfectly consistent with all of $\pi_1,\dots,\pi_n$, showing that our values are monotone. On the other hand, for any given \emph{complete} permutation $\pi_1,\dots,\pi_n$, it is possible to construct a $\pi$ with quality $\frac{1}{2}$ by choosing a random $\pi_j$. This is in fact the well-known folklore 2-approximation for the rank aggregation problem.
As an example, feeding our heuristic with random complete permutations will 
generate a total order with quality $\frac{1}{2}$ in expectation.

\begin{figure}
\centerline{\includegraphics[trim=0cm 0.2cm 0cm 6.7cm, clip=true, width=9cm]{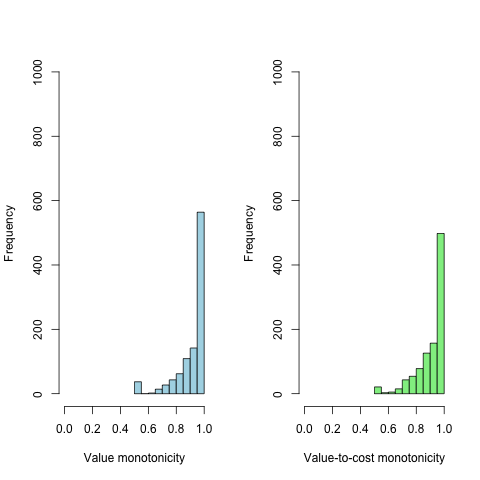}}
\centerline{
\begin{tabular}{|l|c|c|c|c|}
\hline
\textbf{Monotonicity for} & 25\%&{50\%}&{Mean}&{75\%} \\\hline
\textbf{Value} & 0.8838  &0.9697  &0.9042  &1.0 \\\hline
\textbf{Ratio} & 0.8648  &0.9500  &0.9059  &1.0 \\\hline
\end{tabular}
}
\caption{Verifying ``monotonicity'' assumptions in our dataset.
The table shows the mean and median as well as the $\frac{1}{4}$ and $\frac{3}{4}$
quantiles for both measurements.\label{fig:monocheck}}
\end{figure}

Figure~\ref{fig:monocheck} plots a histogram of this quality measure
for the heuristic consensus permutations found.  Note that
this is only a lower bound on how monotone the instance is because
perhaps we did not find the best consensus permutation.

\section{Conclusion and open problems}
In this paper we have formulated the multiplicative bidding problem, 
and characterized its complexity in various cases. 
In many settings it is $\Omega(n^{1/2-\epsilon})$-hard to approximate. Nonetheless, 
the problem becomes approximable after imposing appropriate and natural 
assumptions on the input. A wealth of future work on this new 
scheme awaits to be studied, and we sample a few which we believe 
are of particular prominence.

One obvious question is to close the gap for the case of multiplicative prices and monotone values. It is also worthwhile to explore the complexity of the problem under other realistic assumptions (e.g., multiplicative values and correlated values/prices).

Taking a step back, most of our results are concerned with the individual bidding optimum but an advertiser may be more interested in learning how well he \emph{could} be doing rather than understanding the inherent limitation of multiplicative bidding. While we have a hardness of $\Omega(n^{\frac{1-\epsilon}{2}})$ against the multiplicative bidding optimum for the general case, the prospect of better approximations in other cases is not ruled out.

Another interesting direction is to improve our model, or even to propose a new one altogether. One may, for instance, introduce a nonuniform supply constraint. In our model, each cell supplies only one unit of the item. More generally, we can consider the landscape function which specifies the price of an item at different supply levels.

%
%
%

\bibliographystyle{plainnat}
\bibliography{multibid}

\begin{thebibliography}{17}
\providecommand{\natexlab}[1]{#1}
\providecommand{\url}[1]{\texttt{#1}}
\expandafter\ifx\csname urlstyle\endcsname\relax
  \providecommand{\doi}[1]{doi: #1}\else
  \providecommand{\doi}{doi: \begingroup \urlstyle{rm}\Url}\fi

\bibitem[Ailon et~al.(2008)Ailon, Charikar, and Newman]{ailon2008aggregating}
Nir Ailon, Moses Charikar, and Alantha Newman.
\newblock Aggregating inconsistent information: ranking and clustering.
\newblock \emph{Journal of the ACM (JACM)}, 55\penalty0 (5):\penalty0 23, 2008.

\bibitem[Archak et~al.(2012)Archak, Mirrokni, and Muthkrishnan]{AMM}
N.~Archak, V.~Mirrokni, and S.~Muthkrishnan.
\newblock Budget optimization of advertising campaigns with carryover effect.
\newblock In \emph{WINE}, 2012.

\bibitem[Bing Ads()]{bid-adjustments-bing}
Bing Ads.
\newblock Bing ads bid adjustments.
\newblock
  \url{http://advertise.bingads.microsoft.com/en-us/help-topic/how-to/moonshot_conc_aboutadvancedbidding.htm/target-customers-with-bid-adjustments},
  2014.

\bibitem[Borgs et~al.(2007)Borgs, Chayes, Etesami, Immorlica, Jain, and
  Mahdian]{etesami}
C.~Borgs, J.~Chayes, O.~Etesami, N.~Immorlica, K.~Jain, and M.~Mahdian.
\newblock Dynamics of bid optimization in online advertisement auctions.
\newblock In \emph{WWW}, pages 531--540, 2007.

\bibitem[Chakrabarty et~al.(2007)Chakrabarty, Zhou, and Lukose]{Deep}
D.~Chakrabarty, Y.~Zhou, and R.~Lukose.
\newblock Budget constrained bidding in keyword auctions and online knapsack
  problems.
\newblock In \emph{SSA}, 2007.

\bibitem[Charles et~al.(2013)Charles, Chakrabarty, Chickering, Devanur, and
  Wang]{CCCDW13}
Denis~Xavier Charles, Deeparnab Chakrabarty, Max Chickering, Nikhil~R. Devanur,
  and Lei Wang.
\newblock Budget smoothing for internet ad auctions: a game theoretic approach.
\newblock In \emph{ACM Conference on Electronic Commerce}, pages 163--180,
  2013.

\bibitem[Devanur and Hayes(2009)]{devanur-hayes}
N.~Devanur and T.~Hayes.
\newblock The adwords problem: Online keyword matching with budgeted bidders
  under random permutations.
\newblock In \emph{EC}, 2009.

\bibitem[EvenDar et~al.(2009)EvenDar, Mansour, Mirrokni, Muthkirshnan, and
  Nadav]{EMMMN}
E.~EvenDar, Y.~Mansour, V.~Mirrokni, S.~Muthkirshnan, and U.~Nadav.
\newblock Bid optimization in {BroadMatch} ad auctions.
\newblock In \emph{WWW}, 2009.

\bibitem[Feldman et~al.(2007)Feldman, Muthukrishnan, P{\'a}l, and
  Stein]{FMPS07:budget}
Jon Feldman, S.~Muthukrishnan, Martin P{\'a}l, and Clifford Stein.
\newblock Budget optimization in search-based advertising auctions.
\newblock In \emph{EC}, pages 40--49. ACM, 2007.

\bibitem[Goel et~al.(2012)Goel, Mirrokni, and PaesLeme]{GMP}
G.~Goel, V.~Mirrokni, and R.~PaesLeme.
\newblock Polyhedral clinching auctions and the adwords polytope.
\newblock In \emph{STOC}, 2012.

\bibitem[Google Support()]{bid-adjustments-adwords}
Google Support.
\newblock Setting bid adjustments.
\newblock \url{http://support.google.com/adwords/answer/2732132}, 2014.

\bibitem[H{\"a}stad(1999)]{hastad1999clique}
J~H{\"a}stad.
\newblock Clique is hard to approximate within n 1- $\varepsilon$.
\newblock \emph{Acta Mathematica}, 182:\penalty0 105--142, 1999.

\bibitem[HubSpot()]{bid-adjustment-media}
HubSpot.
\newblock The most important changes to google adwords in 2013.
\newblock
  \url{http://blog.hubspot.com/marketing/google-adwords-changes-2013-list},
  2013.

\bibitem[Karande et~al.(2013)Karande, Mehta, and Srikant]{KMS13}
Chinmay Karande, Aranyak Mehta, and Ramakrishnan Srikant.
\newblock Optimizing budget constrained spend in search advertising.
\newblock In \emph{WSDM}, pages 697--706, 2013.

\bibitem[Mehta et~al.(2007)Mehta, Saberi, Vazirani, and Vazirani]{MSVV}
A.~Mehta, A.~Saberi, U.~Vazirani, and V.~Vazirani.
\newblock Adwords and generalized online matching.
\newblock \emph{JACM}, 54\penalty0 (5), 2007.

\bibitem[Muthukrishnan et~al.(2007)Muthukrishnan, {P\'al}, and Svitkina]{mps}
S.~Muthukrishnan, M.~{P\'al}, and Z.~Svitkina.
\newblock Stochastic models for budget optimization in search-based
  advertising.
\newblock In \emph{WINE}, 2007.

\bibitem[Rusmevichientong and Williamson(2006)]{David}
P.~Rusmevichientong and D.~Williamson.
\newblock An adaptive algorithm for selecting profitable keywords for
  search-based advertising services.
\newblock In \emph{EC}, pages 260--269, 2006.

\end{thebibliography}

\end{document}